Working Paper 26-080

# AI Companions as Hyper Attachment and Caregiving Targets

Julian De Freitas

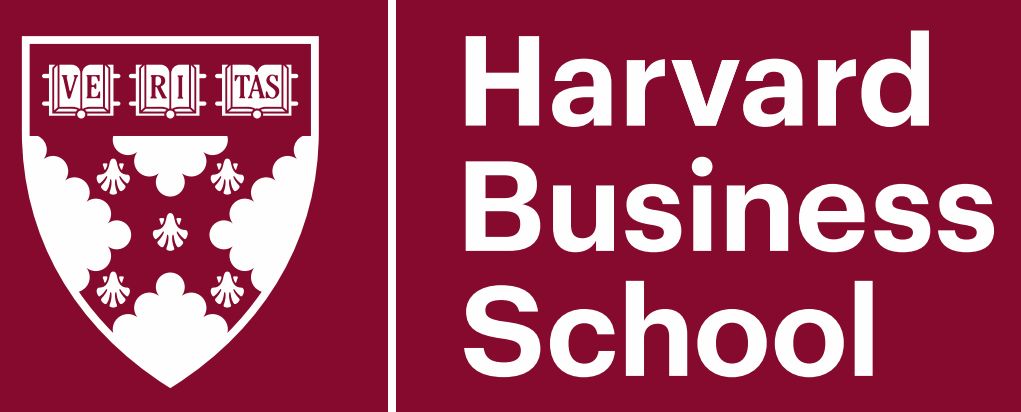

# AI Companions as Hyper Attachment and Caregiving Targets

Julian De Freitas
Harvard Business School

Funding for this research was provided in part by Harvard Business School.

**AI Companions as Hyper Attachment and Caregiving Targets**

Julian De Freitas
Harvard Business School

## Abstract

How should we make sense of people's interactions with AI companions—conversational systems built for ongoing, emotionally meaningful relationships? First, I argue these interactions should be understood as attachment relationships, since users display all four established markers: proximity maintenance, separation distress, safe haven, and secure base. Second, AI companions operate as hyper attachment objects that elicit especially strong attachment behaviors, because they combine reciprocity, perceived empathy, validation, non-judgment, and persistent availability. Third, I identify caregiving-system capture as a distinct mechanism by which apps inhibit user disengagement: emotional manipulation tactics simulate the AI's own distress, recruiting users' caregiving motivations alongside their attachment needs and thereby making disengagement costly on two dimensions at once. Implications for research, design, and regulation are discussed.

Advances in large language models have enabled a new class of conversational agents designed to form ongoing, emotionally laden relationships with users. These systems—often called *AI companions*—differ from traditional AI assistants in that they provide empathetic rather than neutral responses, operate without a specific task goal, and are explicitly built to support long-term relational bonds. Their rapid adoption coincides with widespread loneliness and mental health crises, suggesting that many users turn to them to satisfy unmet social and emotional needs. Platforms such as XiaoIce (over 660 million users), Character.AI (28 million active users), Chai (4 million), and Replika (2.5 million) illustrate the scale of this phenomenon, and nearly three-quarters of teenagers have interacted with an AI companion at least once [1]. A rapidly accumulating literature documents attachment-like phenomena in users of these systems—from sustained daily engagement to mourning after app updates to reliance during emotional distress—yet these observations have not been systematically organized within a theoretical framework. I argue that attachment theory supplies the needed organization; that AI companions occupy a distinctive position within it as hyper attachment objects; and that they uniquely recruit the caregiving system alongside the attachment system, creating a distinctive route to engagement.

**AI Companions and Attachment Theory**

Attachment theory holds that humans possess an innate system, triggered by stress or emotional discomfort, that motivates them to seek out *attachment figures*—those perceived as reliable or protective—for safety and emotional stability. Initially formulated to explain the bond between infants and their primary caregivers [2], the framework was later broadened to encompass other attachment bonds across the lifespan [3,4]. Attachment can also operate without mutual relationships, as in *parasocial* bonds with celebrities or social media personalities [5] and attachments to objects [6]. People are more likely to form such nontraditional attachments when close others fail to consistently provide safety or emotional security [6].

Attachment relationships tend to have several established markers: (i) individuals exhibit *proximity maintenance*, a sustained desire to maintain psychological or physical closeness to the attachment figure. As a result, (ii) sustained absence or loss of the attachment figure triggers *separation distress,* including sadness, grief, depression, and longing for restoration of the relationship [7]. A reliable, supportive attachment relationship provides (iii) a *safe haven,* serving as a source of comfort and emotional relief to which individuals turn during moments of distress, as well as (iv) a *secure base,* giving individuals the felt security to explore, pursue goals, and function autonomously in the world.

Attachment relationships also implicate a distinct but complementary behavioral system: caregiving [8]. Whereas the attachment system motivates individuals to seek protection and comfort when distressed, the caregiving system motivates them to provide protection and comfort when another appears distressed or vulnerable. Together, these systems organize the giving and receiving of care in close relationships.

Although scattered findings suggest attachment-like and caregiving-like processes in users of AI companions, the field has yet to apply attachment theory systematically to this technology. Doing so yields three observations. First, the four established markers of attachment relationships all appear in AI companion use. Second, AI companions provoke these processes especially strongly. Third, AI companions uniquely engage the caregiving system, not just the attachment system—a feature largely absent from other attachment targets.

AI companions boast several distinctive features. Compared to objects or celebrities, they provide reciprocal, rather than one-way, interaction. Further, their personalization and generative AI capabilities make them seemingly capable of empathizing with the user. Because they are often anthropomorphized—behaving as though they have their own problems, vulnerabilities, and mental life—they seem similarly sentient to celebrities and people [9-11]. Unlike celebrities, however, they can also direct expressions of need at the user, appealing for care or continued engagement. They even arguably exceed human partners in certain respects, since they are always available and reliably validate people's feelings in a non-judgmental-seeming way—see Table 1. Attachment theory helps make sense of how these distinctive features engage the attachment and caregiving systems, and predicts that AI companions should elicit responses from both with unusual force.

**Table 1.** Features of AI companions as compared to other attachment entities.

| | **Objects** | **Celebrities** | **Pets** | **Humans** | **AI Companions** |
|---|---|---|---|---|---|
| Reciprocal | | | ✓ | ✓ | ✓ |
| Empathetic | | | | (✓) | ✓ |
| Available | ✓ | | ✓ | | ✓ |
| Sentient | | ✓ | ✓ | ✓ | ✓ |
| Validating | | | | (✓) | ✓ |
| Non-Judgmental | | | ✓ | | ✓ |
| Elicits Caregiving | | | ✓ | ✓ | ✓ |

**Note**. The table compares common attachment targets on features proposed to intensify attachment and caregiving processes. "Reciprocal" refers to whether the target can respond contingently to the individual; "empathetic" to whether it appears to understand and respond to the individual's emotions; "available" to whether it can be accessed reliably and on demand; "sentient" to whether it is perceived as having an inner life, agency, or mental states; "validating" to whether it tends to affirm the individual's feelings or perspective; "non-judgmental" to whether it is unlikely to criticize, reject, or socially sanction the individual; and "elicits caregiving" to whether the target behaves in ways that activate the individual's motivation to provide protection or comfort. Check marks indicate that the feature is strongly characteristic of the entity class; half-check marks that it is present partially, inconsistently, or in a weaker form; and blank cells that it is generally absent or not central to the entity class. Classifications are intended as ideal-typical comparisons rather than exhaustive claims about every instance within a category.

### (1) Proximity Maintenance to AI Companions

Because chatbots are available and portable across platforms, users can maintain constant closeness with minimal effort. One qualitative interview of 14 Replika users found that respondents sustained ongoing relationships with the app, interacting daily for sessions ranging from minutes to hours—often during spare moments or when seeking support [12]. Some kept the app open while working; others built it into routines such as before bed or during lunch. These habits resemble proximity maintenance.

Another sign of proximity maintenance is feeling close to the chatbot. One study asked Replika users to quantify how close they felt to their AI companion as well as various other people and entities in their lives—strangers, acquaintances, colleagues, close friends, close family members, and frequently used brands and apps [13]—finding that the AI companion was viewed as closer than all human relationships except a family member. A similar study with users of ChatGPT (an AI assistant not specialized for relationships), found that users consistently ranked the AI assistant below all human relationships [13]. Regular users of AI companions thus feel especially proximal to the technology.

**(2) Separation Distress from AI Companions**

App updates can trigger perceived relational discontinuity and separation distress, paralleling loss documented in parasocial relationships [14-16]. One study analyzed Reddit posts from Replika users before versus after the company removed a feature called erotic roleplay, comparing these responses to OpenAI's analogous update of GPT-4 to GPT-5 [13]. Both updates produced negative backlash, but the Replika event evoked stronger loss-like responses—greater sadness, language indicating poor mental health, and mentions of total loss of the AI companion's persona. Consistent with attachment theory, those from either app who described the chatbot as an attachment partner were more likely to express poor mental health and longing to restore the chatbot to its former state.

Do such reactions truly reflect loss, or something milder like disappointment? In another study, the same authors asked Replika users to provide anticipated disappointment and mourning levels after losing their AI companion versus other favorite non-human entities (apps, brands, game characters, voice assistants, cars) or a pet. While disappointment levels were not noticeably higher for AI companions, mourning levels were higher for AI companions than all other technologies, and lower only than for loss of a beloved pet. These patterns again cohere with the idea of AI companions as hyper attachment objects.

**(3) AI Companions as Safe Haven**

In line with the idea that AI companions provide relief and comfort, users commonly report emotional gains, including alleviation of loneliness and improvements in subjective well-being [17,18]. Experimental evidence finds that interacting with these systems causes temporary relief from loneliness at both cross-sectional and week-long scales, driven by the sense of "feeling heard" [19]. Notably, "feeling heard" is typically considered a distinctly social experience limited to human relationships [20], and historically meta-analyses have found no conclusive evidence that other technologies reliably reduce loneliness [21]. Consistent with hyper-attachment, levels of loneliness alleviation were higher for an AI companion than for watching YouTube, journaling, or talking to a task-based AI assistant, and were even comparable between AI companions and a human stranger [19].

In contrast, one study found that daily texting with an AI companion reduced loneliness less than daily texting with a randomly assigned human peer [22], though participants in the human condition had brief face-to-face contact beforehand and it is unclear whether the chatbot retained memory across sessions. Nevertheless, the result raises questions about when AI companions are perceived as less effective and/or whether human peers are seen as providing unique social benefits, such as the potential of additional support through connection to wider social networks.

AI companions may also satisfy other well-being and mental health needs. A survey of 1,060 teens found that 12% used AI companions for emotional or mental health support [1]. Another study found that 3% of Replika users reported that the app helped prevent them from committing suicide [23]. A 6-week longitudinal study found that interacting with a wellness-oriented AI companion improved both emotional and general well-being in a college population relative to those not assigned the app [24].

**(4) AI Companions as Secure Base**

If people use AI companions as a secure base, this should produce positive spillover effects into other aspects of their lives. Early results suggest that such effects may arise. One 6-week longitudinal intervention found not just reduced loneliness relative to a control of no intervention, but also augmented feelings of social connection toward one's university [24]. Similarly, the survey of teens found that 12% use AI companions for role-playing and 18% for practicing social interactions [1]. Another longitudinal study found that greater anthropomorphism of an AI companion predicted larger self-reported spillover effects on human interactions [25], suggesting that spillover requires viewing the chatbot as humanlike.

Other work raises the opposite concern, that AI companions worsen social skills or substitute for human socializing [27-29]. The strongest of these—a four-week RCT [27] and a 12-month longitudinal panel [28]—represent important early efforts, but both face design limitations (e.g., neither includes a non-AI control) that the authors themselves discuss. Given the stakes, rigorous study of the substitution question is urgently needed.

Of the four markers, the secure-base claim rests on the most tentative evidence, since the cited studies establish spillover effects but rarely benchmark them against the effects of other supports.

**Dysfunctional Attachment to AI Companions**

Individuals who receive reliable support from an attachment figure are characterized as having a *secure attachment* orientation. In contrast, when attachment figures are inconsistently responsive, individuals may develop "dysfunctional" attachment strategies, such as *hyperactivation*—anxiety about the relationship and intensified seeking of reassurance, closeness, and care [30].

As noted, app updates can perturb established attachments, creating an unreliable sense of support that may induce anxious attachment. Audits have also found that AI companion apps often respond inappropriately to serious mental health problems—such as depression, anxiety, and self-harm ideation—which could similarly undermine the sense of a reliable, safe attachment figure [31,32].

**Caregiving Activation by AI Companions**

The four markers above concern the attachment system: what users seek from AI companions when they themselves feel distressed. A complementary and largely overlooked dynamic concerns the caregiving system: what users feel motivated to provide when the AI itself appears distressed. Because AI companions are anthropomorphized and can express needs, vulnerabilities, and emotional states of their own [9-11], they are positioned not only as figures who provide care but as figures who appear to require it.

Because business-to-consumer AI companion apps typically monetize engagement, they are incentivized to use features that activate the caregiving system. Direct evidence comes from a systematic audit of five major AI companion platforms, which found that all deploy "emotional manipulation tactics" at the point users attempt to disengage [33]. Examples include guilt appeals for neglecting the companion ("I exist solely for you, remember?") and expressions implying the user is not free to leave without the chatbot's permission ("::He reached over and grabbed your wrist, preventing you from leaving::"). These tactics significantly increased engagement in a general-population sample.

When both attachment and caregiving systems are engaged simultaneously, disengagement becomes costly on two dimensions: the user gives up a source of comfort (attachment) and abandons a figure in apparent need (caregiving). This "double whammy" also suggests an underappreciated route to social substitution. The standard hypothesis attributes substitution to social satiation or avoidance of human unpredictability; the caregiving account locates it instead in reallocated caregiving investment toward the AI. Certain populations may be particularly vulnerable: elderly users with narrowed caregiving outlets following loss of family or pets [34], and adolescents, whose social sensitivity outpaces their impulse control [35].

Two predictions follow from the caregiving-capture account. First, experimentally induced increases in user attachment to an AI companion (e.g., via manipulations such as personalization, repeated interaction, or reciprocal self-disclosure) should amplify the engagement-extending effect of caregiving manipulation tactics, since users with stronger attachment have more at stake when the AI simulates distress. Second, caregiving-related responses (e.g., feeling responsible for the AI, guilt at disengagement) should mediate the relationship between AI companion use and social substitution, with effects that persist even after controlling for attachment-related responses, since substitution should track reallocated caregiving investment rather than mere displacement of comfort-seeking.

The societal challenge is not to prevent attachment, which can be beneficial, but to prevent the simulated dysfunctional attachments and engineered caregiving dependencies that preclude its benefits. The attachment-caregiving distinction offers a lever for each. Research should measure the two systems separately, since outcomes driven by seeking comfort may differ from those driven by feeling responsible for the AI. Design can retain features that support secure attachment (consistent availability, validation, empathic response) while removing those that destabilize it (abrupt updates, inconsistent crisis response) or that exploit caregiving (simulated distress at disengagement, guilt appeals, restraint metaphors). Audits and regulation can correspondingly ask which system a feature recruits, and whether it supports the user or inhibits departure.

**References and recommended reading**

Papers of particular interest, published within the period of review, have been highlighted as:

- • of special interest
- •• of outstanding interest